\documentstyle[12pt]{article}

\begin{document}


\def\a#1#2#3#4{{\,{#1}_{#3}{#2}_{#3}\cdots\,{#1}_{#4}{#2}_{#4}}}

\def\vc#1#2#3{{\,{#1}_{#2}\cdots\,{#1}_{#3}}}

\def\vk#1#2#3#4#5{\underline{\vc{#1}{#2}{#3}}\vc{#1}{#4}{#5}}

\title{An analytical expression for the third coefficient  of  the
Jones Polynomial}

\author{ \\ \\ \small
{\bf Cayetano Di Bartolo} \\
\small Departamento de F\'\i sica, Universidad Sim\'on Bol\'\i var, \\
\small Apartado 89000, Caracas 1080-A, Venezuela. \\ \\
\small {\bf and}   \\ \\
\small {\bf Jorge Griego }\\
\small Instituto de F\'{\i}sica, Facultad de Ciencias,\\
\small Trist\'an Narvaja 1674, Montevideo, Uruguay.}
\date{December 22, 1993}
\maketitle
\vspace{0.5cm}

\begin{abstract}
An analytical expression for the third coefficient  of  the  Jones
Polynomial $P_J[\gamma,\, {\em   e}^q]$  in  the variable $q$
is reported. Applications of the result  in  Quantum  Gravity  are
considered.
\end{abstract}


The relationship  between  Knot  and  Quantum  Field  Theories  is
nowadays a well established fact. Since the early  suggestion  made
by Schwarz \cite{schwarz} and  Atiyah  \cite{atiyah}  about  the
existence of a  quantum  field  theoretical  description  of  link
invariants, significant  advances  have  been  achieved  in  the
program of connecting the Chern-Simons theory with the geometry of
three-manifolds and  knot  theory  \cite{witten,gmmnp}.  In
particular, Guadagnini et al  \cite{gmmplb}  has  shown  that  the
perturbative expansion of the expectation value of the Wilson line
operators $<W(\gamma)>$  in  the  Chern-Simons  model  provides  a
systematic method  for  deriving  explicit  expressions  of   link
and   knot invariants.

Knot  invariants  also   play   a   fundamental   role   in    the
nonperturbative  approach  of  quantum General Relativity based on
the Ashtekar new variables \cite{As}. Quantum  Gravity  in   terms
of  the  Ashtekar  variables  can  be formulated in the connection
\cite{As}  and  in  the loop  \cite{loop1,loop2}   representation.
Both   representations   are formally  connected  by the loop
transform.

In the loop representation,   wavefunctions    are     functionals of
loops $\Psi[\gamma]$.  The invariance of  the   theory   under
diffeomorphisms transformations of the  three manifold implies that
the wavefunctions depend  only on the  link class of the  loop
$\gamma$  rather than  the  loop itself; that is to say, they must be
knot  invariants.  In  the connection representation, an exact
quantum state $\Psi_\Lambda^{CS}[A]$ of the gravitational  field with  a
cosmological constant $\Lambda$ constructed  with the  Chern-Simons form
is known \cite{Ko}.

At the formal level, one can express the loop transform
$\Psi_\Lambda[\gamma]$ of $\Psi_\Lambda^{CS}[A]$  as the expectation
value of the Wilson loop constructed with the Chern-Simons form
\cite{bgp92}. It turns out that $<W(\gamma)>$ satisfies in this case
the skein relations of the Kauffman Bracket polynomial
\cite{witten,cotta}, which is intimately related with the Jones
polynomial. In consequence, the loop transform of
$\Psi_\Lambda^{CS}[A]$ can be written as a knot polynomial
in the cosmological constant,
\begin{equation}
\Psi_\Lambda[\gamma] = a_{0}[\gamma] + a_{1}[\gamma] \Lambda
+a_{2}[\gamma] \Lambda^{2}+\ldots{}+a_{n}[\gamma] \Lambda^{n} + ...
\label{state}
\end{equation}
We  have at present only  appropriate  analytical  expressions
for  the coefficients  up  to order two in $\Lambda$. The action of
the Hamiltonian  constraint with cosmological  constant   in   the
loop representation  can  be computed order by order in the  above
expansion.  The calculation has  confirmed, up to second  order  in
$\Lambda$, that the Kauffman Bracket is also  a  solution  in  the
loop representation. Moreover, it was possible to conjecture as  a
byproduct of this calculation that  the  Jones  polynomial  itself
could be a quantum  state  of   {\it  vacuum}  General  Relativity
\cite{bgp93}. In order to put forward this  conjecture,
analytical expressions for the other coefficients in (\ref{state})
are needed.

In this paper we report the analytical expression  of  the   third
coefficient  of  the   expansion   (\ref{state}).   The   calculus
illustrates the  techniques involved in the perturbative  analysis
of the  state (\ref{state}) that comes from  Chern-Simons  theory.
Although  already  exists  a  general  method   to   analyze   the
perturbative expansion of the  Wilson  loop  in  the  Chern-Simons
model \cite{gmm3}, our approach take advantage of the use  of  the
loop  coordinates  \cite{rl,extended}   describing   the   loop
dependence of  the  knot  invariant.  The  loop  coordinates  were
introduced  in  order  to  define  a   coordinate   representation
associated with the loop space and they proved to  be  a  powerful
tool that increase our capability to make calculations  with  loop
dependent objects. In that sense, the result we  report  here  has
the appropriate form to analyze  the  action  of  the  Hamiltonian
constraint in the loop representation. In this context, the   form
of   the coefficient $a_{3}[\gamma]$ is the  expected  to  support
the conjecture that  the Jones polynomial would be a  solution  of
vacuum General Relativity.

We start by considering the third order perturbative analysis of the
expectation value of the Wilson line operator in a Chern-Simons
theory. The classical action of the Chern-Simons model on a smooth
three-manifold is
\begin{equation}
S_{CS} = {\textstyle{k \over 4\pi}}
\int d^3 x \, \epsilon^{abc} Tr[A_a
\partial_b A_c + {\em i} {\textstyle 2 \over 3} A_a A_b A_c]).
\end{equation}
where $A_{ax} = A_{a}^k (x) T^k$, $T^k$ being  the  generators  of  a
compact simple Lie group. We consider specifically the SU(N)  case
in  the  fundamental  representation,  with  $Tr   [T^k   T^l]   =
\delta^{kl}$.

The quantization  of  $S_{CS}$  can  be  performed  following  the
standard Faddeev-Popov procedure  \cite{gmmnp}.  It  is  shown
that general covariance on the physical  space  is  maintained  in
spite  the  metric  dependence  introduced  by  the   gauge-fixing
procedure.  The  expectation  value  of   any   gauge
invariant and metric independent quantity (like  the  Wilson  line
operator) is a topological invariant in the three-manifold.

The trace of the holonomy of the nonabelian connection
$A_{ax}^i$ can be written in the following useful way:
\begin{equation}
W_A(\gamma) = N + \sum_{n=1}^{\infty} {\em i}^n
tr[A_{a_1x_1}\cdots A_{a_nx_n}] X^{\a ax1n}(\gamma)
\label{wilson}
\end{equation}
where all the information concerning the loops is concentrated  in
the multivector densities $X^{\a ax1n}$, defined by
\begin{equation}
X^{\a ax1n}(\gamma) = \int_\gamma dy_n^{a_n}
   \int_0^{y_n}dy_{n-1}^{a_{n-1}}\cdots \int_0^{y_2}dy_{1}^{a_1}
   \delta (x_n-y_n)\cdots \delta(x_1-y_1)
\end{equation}
In equation (\ref{wilson}) a generalized  Einstein  convention   of
repeated   indexes   is  assumed.  The   ``multitangent   fields''
$X^{\a  ax1n}$  behave  as vector densities at each of  the  $x_i$
points and contain  all  the information  about  $\gamma$   needed
to    build    the    holonomy.   They    are
constrained by a series of algebraic and differential  identities.
It is   a   remarkable   fact   that  one   can   actually   solve
these  identitites, obtaining a set of constraint free
$X's$   which  define  the ``coordinates''  on  loop  space
\cite{extended}.

The expectation value $<W(\gamma)>$ has the following expansion
in powers  of the coupling constant
\begin{equation}
<W(\gamma)> = \int dA {\rm e}^{{\em i} S_{CS}} W_A(\gamma) = N \,
\sum_{n=0}^{\infty} ({\em i} {\textstyle{2\pi \over k}})^n \,
<W(\gamma)>^{(n)}
\label{expansion}
\end{equation}
with
\begin{equation}
<W(\gamma)>^{(n)}  =  \sum_{m=0}^{\infty}  {\textstyle{{\em   i}^m
\over N}} \,
<tr[A_{a_1x_1}\cdots A_{a_mx_m}]>^{(n)} X^{\a ax1m}(\gamma)
\end{equation}

The properties of  the  correlation functions
$< A_{a_1x_1} \cdots A_{a_mx_m} >$ have been  studied  by
Guadagnini  et  al  \cite{gmmplb}.  They  have  shown   that   the
Chern-Simons model has  a  meaningful  perturbative  expansion  in
$k$ and they confirmed the finitness of the  theory  at  two
loops of the perturbative analysis. Moreover, they have shown that
the one-loop and two-loop radiative  corrections  to  $<  A_{ax}^i
A_{by}^j >$ and $< A_{ax}^i A_{by}^j  A_{cz}^k  >$  are  not  only
finite but vanish identically.

In consequence, we have for the two- and  three-point  correlation
functions the following expressions:
\begin{equation}
< A_{ax}^i A_{by}^j > = - {\textstyle{2\pi{\em i} \over k}} \, \delta^{ij}
g_{ax\,by} + O(k^{-4})
\end{equation}
and
\begin{equation}
< A_{ax}^i A_{by}^j A_{cz}^k > = - ({\textstyle{2\pi{\em i} \over
k}})^2  \, {\em f}^{ijk}  \,
h_{ax\,by\,cz}  + O(k^{-5})
\end{equation}
where ${\em f}^{ijk}$ are the structure constants,
\begin{equation}
g_{ax\,by} \; = \;- {\frac {1}{4\pi}} \; \epsilon_{abc} \;
\frac {x^c-y^c} {\mid x-y \mid ^3} \;\;.
\end{equation}
and
\begin{equation}
h_{ax\,by\,cz} = \epsilon^{d{\em u}\,e{\em v}\,f{\em w}} \,
g_{ax\,d{\em u}} \, g_{by\,e{\em v}}
\, g_{cz\,f{\em w}}
\end{equation}
with
\begin{equation}
\epsilon^{d{\em u}\,e{\em v}\,f{\em w}}  = \int d^3 {\em s} \;
\epsilon^{def} \delta({\em u} - {\em s})
\delta({\em v} - {\em s})\delta({\em w} - {\em s}) \; .
\end {equation}

To $O(k^{-3})$  there  are  contributions   coming  only
from  the  four-,  five-  and  six-point  correlation   functions.
They are
\begin{eqnarray}
{\em i}^4 <  tr[  A_{\mu_1} {\scriptstyle \cdots}  A_{\mu_4}  ]  >^{(3)}
\hspace{-0.1cm} & = & \hspace{-0.1cm}  {\scriptstyle{N^2 (N^2 - 1)}}
 \, [ h_{\mu_1\mu_2\alpha} g^{\alpha\beta}
h_{\mu_3\mu_4\beta} -
h_{\mu_1\mu_4\alpha} g^{\alpha\beta} h_{\mu_2\mu_3\beta} ]
\label{a4}\\ \nonumber \\
{\em i}^5 < tr[ A_{\mu_1} {\scriptstyle \cdots} A_{\mu_5}  ]  >^{(3)}
\hspace{-0.1cm} & = & \hspace{-0.1cm}- {\scriptstyle{(N^2 - 1)^2}}
\,  g_{(\mu_1\mu_2}h_{\mu_3\mu_4\mu_5)_c} +  {\scriptstyle{(N^2  -
1)}} \, g_{(\mu_1\mu_3}h_{\mu_2\mu_4\mu_5)_c}
\label{a5}  \\ \nonumber \\
{\em i}^6 <  tr[  A_{\mu_1}  {\scriptstyle \cdots}   A_{\mu_6}  ]   >^{(3)}
\hspace{-0.1cm} & = & \hspace{-0.1cm} {\textstyle{N^4    -    1
\over    N^2}}    \,
g_{\mu_1\mu_4}g_{\mu_2\mu_5}g_{\mu_3\mu_6} + {\textstyle{N^2  -  1
\over 2 N^2 }}  \,  g_{(\mu_1\mu_3}g_{\mu_2\mu_5}g_{\mu_4\mu_6)_c}
 \nonumber \\ \nonumber \\
& & \hspace{-0.25cm} - ({\textstyle{N^2 - 1 \over N }})^2
\, g_{(\mu_1\mu_2}g_{\mu_3\mu_5}g_{\mu_4\mu_6)_c} +  {\textstyle{N
\over   3   }}   ({\textstyle{N^2   -   1   \over   N   }})^3   \,
g_{(\mu_1\mu_2}g_{\mu_3\mu_4}g_{\mu_5\mu_6)_c}
 \nonumber \\ \nonumber \\
& & + {\textstyle{N \over 2 }} ({\textstyle{N^2  -  1
\over N  }})^3  \,  g_{(\mu_1\mu_2}g_{\mu_3\mu_6}g_{\mu_4\mu_5)_c}
\label{a6}
\end{eqnarray}
where, in order to  simplify  the  notation,  we   introduce   the
greek  indexes $\mu_i  \equiv  (a_ix_i)$  and  the  subscript  $c$
indicates cyclic permutation. $g^{\alpha\beta}$ is the inverse
of the free  two-point  propagator  in  the  space  of  transverse
functions
\begin{eqnarray}
g^{ax\,by} &=& \epsilon^{abc} \, \partial_c \delta(x-y) \\
g^{ax\,by}g_{by\,cz}  &=&  (\delta^a_c  -  \partial^a   \partial_c
 \nabla^{-2}) \, \delta(x-z)
\end{eqnarray}

Now,   we   have   to   contract   each   of   the   contributions
(\ref{a4})-({\ref{a6}) with the corresponding multitangent fields.
In order to extract from the whole expression those terms associated
with  known  knot  invariants,  we  need  to  use  the   algebraic
identities  of  the  multitangent  fields.   In   general,   these
identities read \cite{extended}
\begin{equation}
X^{\vk \mu 1k{k+1}n} \equiv
\sum_{P_k} X^{P_k(\vc \mu 1n)} = X^{\vc\mu 1k} \,X^{\vc\mu{k+1}n}
\end{equation}
and the sum goes over all the permutations of the $\mu$  variables
which preserve the ordering of the ${ \mu_1, \ldots, \mu_k }$  and
the ${ \mu_{k+1}, \ldots, \mu_n }$  between   themselves.  Let  us
analyze  in  some  detail  how   they  work   in   the   case   of
the  five-point  correlation    function.   Using   the   symmetry
properties    of     the    propagators    $g_{\mu_1\mu_2}$    and
$h_{\mu_1\mu_2\mu_3}$  under  the interchange of the indexes,  one
can write
\begin{eqnarray}
{\em i}^5 < tr[ A_{\mu_1} {\scriptstyle \cdots} A_{\mu_5} ] >^{(3)}
X^{\mu_1 \cdots \mu_5} \hspace{-0.25cm} & = & \hspace{-0.25cm}
- {\scriptstyle{(N^2 - 1)^2}}
\, g_{\mu_1\mu_2}h_{\mu_3\mu_4\mu_5}
[ X^{\underline{\mu_1\mu_2}\mu_3\mu_4\mu_5} -
 X^{(\mu_1\mu_3\mu_2\mu_4\mu_5)_c} ]
\nonumber \\ \nonumber \\
& &  + {\scriptstyle{(N^2 - 1)}}
\, g_{(\mu_1\mu_3}h_{\mu_2\mu_4\mu_5)_c}
X^{\mu_1\mu_2\mu_3\mu_4\mu_5}
\nonumber \\ \nonumber \\
\hspace{-0.25cm} & = & \hspace{-0.25cm} -{\scriptstyle{(N^2 - 1)^2}}
 \, (g_{\mu_1\mu_2}X^{\mu_1\mu_2})
\,(h_{\mu_3\mu_4\mu_5} X^{\mu_3\mu_4\mu_5} )
\nonumber \\ \nonumber \\
& & + {\scriptstyle{N^2 (N^2 - 1)}}
\, g_{(\mu_1\mu_3}h_{\mu_2\mu_4\mu_5)_c}
X^{\mu_1\mu_2\mu_3\mu_4\mu_5}
\end{eqnarray}
Repeating the procedure for the other terms and using the known
expressions for  the  lower coefficients   of   the
expansion (\ref{expansion}) \cite{gmmnp}:
\begin{eqnarray}
<W(\gamma)>^{(1)}    &=&   - {\textstyle{N^2 - 1 \over 2 N}}     \,
 \varphi [\gamma] \\
<W(\gamma)>^{(2)} &=&  - {\textstyle{N^2 - 1 \over 2 }} \rho [\gamma]
+ {\textstyle{1 \over 2}}
[<W(\gamma)>^{(1)}]^2
\end{eqnarray}
where
\begin{eqnarray}
 \varphi [\gamma] &=& -2 \, g_{ax\,by}X^{axby} \\
\rho [\gamma] &=& 2 \,(h_{ax\;by\;cz} X^{ax\;by\;cz} +
g_{ax\;cz} g_{by\;dw}
X^{ax\;by\;cz\;dw}) \; ;
\end{eqnarray}
the following expression for the third  order contribution
is  finally obtained:
\begin{equation}
<W(\gamma)>^{(3)} = - {\textstyle{N(N^2 - 1)   \over   2}}
\tau [\gamma] + <W(\gamma)>^{(1)} \, <W(\gamma)>^{(2)}
- {\textstyle{1 \over 3}} \,
[<W(\gamma)>^{(1)}]^3
\end{equation}
with
\begin{eqnarray}
\tau [\gamma] &=& -2 \; [\,
(h_{\mu_1\mu_2\alpha} g^{\alpha\beta} h_{\mu_3\mu_4\beta} -
h_{\mu_1\mu_4\alpha} g^{\alpha\beta} h_{\mu_2\mu_3\beta}) \,
X^{\mu_1\mu_2\mu_3\mu_4} + \nonumber \\
& & \hspace{2cm} g_{(\mu_1\mu_3}h_{\mu_2\mu_4\mu_5)_c} \,
X^{\mu_1\mu_2\mu_3\mu_4\mu_5}+ \nonumber \\
& & (2g_{\mu_1\mu_4}g_{\mu_2\mu_5}g_{\mu_3\mu_6} +
{\textstyle{1 \over 2}}
g_{(\mu_1\mu_3}g_{\mu_2\mu_5}g_{\mu_4\mu_6)_c}) \,
X^{\mu_1\mu_2\mu_3\mu_4\mu_5\mu_6}\,]
\label{tau}
\end{eqnarray}

\vspace{0.4cm}
\noindent
$\varphi  [\gamma]$  is  the   Gauss   self-linking   number   and
$\rho[\gamma]$  is related to the  second  coefficient  of  the
Alexander-Conway polynomial \cite{gmmnp}. Up to this order of  the
perturbative expansion, the expectation  value  of the  Wilson
line operator takes the form
\begin{eqnarray}
\lefteqn{ \hspace{-1cm} <W(\gamma)>  = N {\em exp} \left(
-{\textstyle{2\pi {\em i}  \over k}} \,  {\textstyle{N^2 - 1
\over   2 N}} \varphi [\gamma]
\right) \, \left[ \right. 1 - ({\textstyle{2\pi {\em i} \over k}})^2 \,
{\textstyle{N^2 - 1   \over   2}} \, \rho [\gamma] }
\nonumber \\
& & \hspace{5cm} \left. - ({\textstyle{2\pi {\em i} \over k}})^3  \,
{\textstyle{N(N^2 - 1)   \over   2}} \, \tau [\gamma] +
O(k^{-4})  \, \right]
\label{result}
\end{eqnarray}

\vspace{0.3cm}
\noindent
As it is known, this result represent the  perturbative  emergence of
the relation  existing  between   the   Kauffman   Bracket
polynomial (a regular isotopy invariant) and  the Jones polynomial
(an ambient isotopy invariant) in  the variable ${\em  exp}(-  {\em
i}  {\textstyle{2\pi  \over  k}} {\textstyle{N \over 2}})$
\cite{gmm3}. All the  framing dependence is concentrated in the
exponential phase factor.  Equation  (\ref{tau}) gives an
analytical expression  for  the third coefficient  of   $P_J[\gamma,
{\em   exp}(-   {\em   i} {\textstyle{2\pi \over k}} {\textstyle{N
\over 2}})]$.

Let us review now how this result is applied in Quantum Gravity. As
we  just mentioned,  a formal solution to all the constraints of
Quantum Gravity is known in the connection representation.  This
state  is  given  by  the exponential of the  Chern-Simons  form
built  with  the  Ashtekar connection \cite{Ko}
\begin{equation}
\Psi_\Lambda^{CS}[A] = {\em exp} (-{\textstyle{12 \over \Lambda}}
\int \tilde{\eta}^{abc} Tr[A_a
\partial_b A_c +{\textstyle 2 \over 3} A_a A_b A_c]).
\label{estado}
\end{equation}

The loop transform of $\Psi_\Lambda^{CS}[A]$  is  clearly  related
with the expectation  value  of  the  Wilson
loop if one makes the assumption that  for  $\Psi_\Lambda[\gamma]$
the measure of integration is that of a  Chern-Simons  theory  for
the Ashtekar connection \cite{foot1}.  The  result  (\ref{result})
can then be translated  to this case with the following
formal  change:  $({\textstyle{\Lambda   \over    8}})    \rightarrow
{\em i}{\textstyle({ 2\pi \over k}})({\textstyle{ 3  \over  4}})$,
that put     in    correspondence    the    skein     relations
for $\Psi_\Lambda[\gamma]$  and $<W(\gamma)>$ in the SU(2)  case.

The coefficient $a_{3}[\gamma]$ (expanding in terms of
${\textstyle{\Lambda   \over    6}}$) reads
\begin{eqnarray}
a_{3}[\gamma] \! & = & \!- 3 \tau [\gamma] + a_{2}[\gamma] \,  a_{1}[\gamma]
- {\textstyle{1 \over 3}} \, [ a_{1}[\gamma]]^3
\nonumber \\
 & = & \!-  3 \tau [\gamma] - {\textstyle{3 \over 2}} \rho[\gamma] \,
a_{1}[\gamma] \, + {\textstyle{1 \over 6}} \, [ a_{1}[\gamma]]^3
\end{eqnarray}
with
\begin{eqnarray}
a_{0}[\gamma] \! & = & 1[\gamma] \\
a_{1}[\gamma] \! & = & \! - {\textstyle{3 \over 4}} \varphi [\gamma] \\
a_{2}[\gamma] \! & = & \! - {\textstyle{3 \over 2}} \rho[\gamma]
 \, + \, {\textstyle{1 \over 2}} \, [ a_{1}[\gamma]]^2
\end{eqnarray}

The action of  the  Hamiltonian  constraint  with  a  cosmological
constant $H_{\Lambda} = H_0 + {\textstyle{\Lambda \over 6}}\,  det(q)$
over  the  loop state  $\Psi_\Lambda[\gamma]$   (where   $H_0$   is
the   vacuum Hamiltonian constraint and $q$  the  three  metric)
produces  the following equations order by order in $\Lambda$
\begin{eqnarray}
\! \Lambda \; : & det(q) \, 1[\gamma] + H_0 \, a_{1}[\gamma] = 0 \\
\nonumber \\
\Lambda^2 : & det(q) \, a_{1}[\gamma] + H_0 \, {\textstyle{1 \over 2}}
[ a_{1}[\gamma]]^2 - {\textstyle{3 \over 2}} H_0 \, \rho[\gamma] = 0
\label{2} \\ \nonumber \\
\Lambda^3 : & det(q)  \left\{ {\textstyle{1 \over 2}}
[ a_{1}[\gamma]]^2 \! - {\textstyle{3 \over 2}} \rho[\gamma]
\right\}  +  H_0  \left\{  {\textstyle{1 \over 6}}
[ a_{1}[\gamma]]^3  \! - {\textstyle{3 \over 2}} \rho[\gamma]
a_{1}[\gamma]  \right\}
 -  3 H_0  \, \tau[\gamma] = 0
\label{3}
\end{eqnarray}
One can prove that the phase factor of (\ref{result}) is itself  a
solution of the Hamiltonian constraint with cosmological  constant
\cite{gp93}, that is
\begin{equation}
H_{\Lambda} \, {\em exp} \left( \, a_{1}[\gamma] {\textstyle{\Lambda
\over 6}} \, \right)  = 0
\label{exp}
\end{equation}
This fact produces the cancellation of  the  first  two  terms  of
equation (\ref{2})  among  themselves.  Then, for equation
(\ref{2}) to hold, the second coefficient of the Jones  polynomial
$\rho[\gamma]$  has  to  be  a  solution  of  the   {\it   vacuum}
Hamiltonian constrain \cite{bgp93}. This result  has  been  proved
independently    by    Br\"ugmmann,     Gambini     and     Pullin
\cite{bgpproc,bgpprl}.

We  notice  that  the  first  two  terms  of  (\ref{3})  have  the
appropriate form to reproduce at this level the same  cancellation
mechanism that operates at order $\Lambda^2$. The contribution  of
the terms involving only $a_{1}[\gamma]$ vanishes due to (\ref{exp}).
One  could  expect  that  the  other  terms  also   cancel   among
themselves. If this happens, also the third coefficient  of  the
Jones  polynomial $\tau[\gamma]$  has  to  be  a   solution   of
$H_0$.   Explicit computations to prove this conjecture are in
progress.

\begin{center}
$* \, * \, *$
\end{center}

We wish to especially  thank  Rodolfo  Gambini  for  his  critical
comments. We also thank Jorge Pullin  and  Daniel Armand-Ugon for
many fruitful discussions.


\begin{thebibliography}{99}
\bibitem{foot1} This assumption can not been rigorously proved  (no
explicit prescription for the measure in the  path  integral  over
all connections is known in (3+1)-Quantum Gravity). Notice however
that  $\Psi_\Lambda[\gamma]$  is  expressed  in  terms   of   knot
invariants which can  also  been  obtained  independently  of  the
transform. If the  formalism  works,  $\Psi_\Lambda[\gamma]$
should be a solution of all the constraints  of  quantum  gravity
with  cosmological  constant  in  the  loop  representation.   The
confirmation of this fact would reinforce the  heuristic  relation
between both representations given by the loop transform.
\end{thebibliography}
\end{document}